\documentclass[
prd,
amsfonts,
preprint,
nofootinbib,
superscriptaddress,
showpacs,
]{revtex4-2}
\usepackage{graphicx, epsfig}
\usepackage{color}
\usepackage{amsmath}
\usepackage{amssymb}
\usepackage{physics}
\usepackage{lineno}
\modulolinenumbers[5]

\newcommand{\be}{\begin{equation}}
\newcommand{\ee}{\end{equation}}
\newcommand{\bea}{\begin{eqnarray}}
\newcommand{\eea}{\end{eqnarray}}

\newcommand{\gapp}{\mathrel{\raise.3ex\hbox{$>$}\mkern-14mu
\lower0.6ex\hbox{$\sim$}}}
\newcommand{\lapp}{\mathrel{\raise.3ex\hbox{$<$}\mkern-14mu
\lower0.6ex\hbox{$\sim$}}}
\def\bbox{{\,\lower0.9pt\vbox{\hrule \hbox{\vrule height 0.2 cm
\hskip 0.2 cm \vrule  height 0.2 cm}\hrule}\,}}

\usepackage{color,xcolor}

\begin{document}
\title{
Studies on particle creation during the universe expansion with a laser system
 }
\def\FDU{Key Laboratory of Nuclear Physics and Ion-beam Application (MoE), Institute of Modern Physics, Fudan University, Shanghai 200433,  China}

\author{De-Chang Dai}
\email{Corresponding author: diedachung@gmail.com}
\affiliation{  Department of Physics, national Dong Hwa University, Hualien, Taiwan, Republic of China}
\affiliation{  CERCA, Department of Physics, Case Western Reserve University, Cleveland OH 44106-7079}
\author{Changbo Fu}
\affiliation{\FDU}

\begin{abstract}
\widetext

While two highly intensive laser beams collide, they create a region where the refractive index varies so quickly that photons are created. The variance of the refractive index is analog to the universe scale factor variance. Therefore, this laser system can be an analog to the expansion of the universe. We find that several hundreds of photons can be created under feasible conditions. This system can demonstrate the particle creation during inflation or other similar periods. 

\end{abstract}


\maketitle

\section{introduction}
 
A black hole is undoubtedly among the most interesting objects in the universe. Classically, it absorbs everything since nothing can escape from the horizon. However, when quantum mechanics is included, Hawking found that a black hole is a thermal object that radiates energy at a certain temperature\cite{Hawking:1974rv,Hawking:1975vcx}. 
A black hole's temperature is inversely proportional to its mass. A large astrophysical black hole is too massive to have an observable radiation. Because of this, Unruh proposed to use non-gravitational systems to mimic the behavior of the gravitational ones \cite{Unruh:1980cg}. These systems are known as analog black holes. 
 
 At first, such analog systems were constructed using water waves \cite{Rousseaux:2007is,Jannes:2010sa,Weinfurtner:2010nu,Euve:2014aga,Euve:2015vml,Torres:2016iee}, and then the methods were  extended to Bose-Einstein condensates \cite{Lahav:2009wx,Steinhauer:2014dra,Steinhauer:2015saa,MunozdeNova:2018fxv,Syu:2022cws,Syu:2019gqq}
 and optical systems \cite{Philbin:2007ji,Belgiorno:2010wn,2012PhRvL.108y3901R,2015PhRvL.114c6402N,Roger:2016slp,Bekenstein,Drori:2018ivu,2012PhRvA..86f3821E}. Today, analog black holes are found in other systems as well, see for example reviews in \cite{Barcelo:2005fc,Barcelo:2018ynq,Rosenberg:2020jde}. 
 
 Though there are huge successes in constructing analog black holes, registering Hawking radiation is a subtler issue. For example, it was claimed that Hawking radiation has been found in an optical system\cite{Belgiorno:2010wn}, but it has also been pointed out that this radiation was just Cerenkov radiation\cite{Belgiorno:2010zz,DallaPiazza:2012aj}. More effort is required to clarify this issue.  

It is unlikely to reproduce inflation in a laboratory. However, it is possible to create a system analogous to an expanding or shrinking universe. We propose that a collision of two laser beams can be analogous to the universe expansion. Several hundreds of photons are created when two pulse lasers cross each other. This particle creation is also an analog to a Schwinger effect\cite{Schwinger:1951nm}. In the following, we review the optical analog system and then demonstrate that this setup is feasible with the present experimental technology.

 \section{Particle creation in nonlinear dielectric material}

Optical analog black holes are created in nonlinear dielectric systems, where the material polarization responds to the external electric field as 

\begin{equation}
\vec{P}=\epsilon_0 \left[\chi^{(1)}\vec{E} +\chi^{(3)}(\vec{E}\cdot \vec{E}) \vec{E}\right]+\ldots, 
\end{equation}
here, $\vec{E}\cdot \vec{E}$ is proportional to the intensity, $I$, of the laser. The propagation of an electromagnetic wave is given by
\begin{equation}
\nabla \times (\nabla \times \vec{E})+\mu_0 \partial_t^2 \vec{P} +\partial_t^2 \vec{E}=0,
\end{equation}
which depends on its intensity. The speed of light is set to
$c=1$ for convenience. Using $\nabla \times (\nabla \times \vec{E})
\approx -\nabla^2 \vec{E}$,
this equation can be simplified to \cite{Rubino:2011zq}
\begin{equation}
\partial_x^2  \vec{E} +\partial_\perp^2  \vec{E}-\mu_0 \partial_t^2 \vec{P} - \partial_t^2 \vec{E}=0 .
\end{equation}

In general, the refractive index,  $n^2= (1+\chi^{(1)}+\chi^{(3)}|\vec{E_p}|^2)$, is controlled by an intensive pump field, $\vec{E}_p$, which then induces a small probe field, $\vec{E}_s$. If the probe field's frequency is much larger than the variance of the intensity, then the probe field's equation of motion is approximated to   
\begin{equation}
(\partial_x^2+\partial_\perp^2 -n^2\partial_t^2 )\vec{E}_s = 2\chi^{(3)} \partial_t^2 (\vec{E}_p \cdot \vec{E}_s) \vec{E}_p. 
\end{equation}
 The probe field has two polarizations. The equation of motion for the probe field perpendicular to the pump field, $\vec{E}_p \cdot \vec{E}_s =0$, can be further simplified to   
\begin{equation}
\label{eom}
(\partial_x^2+\partial_\perp^2 -n^2 \partial_t^2 )\vec{E}_s = 0 .
\end{equation}

Based on this equation, an analog metric is introduced \cite{Belgiorno:2010iz}
\begin{equation}
\label{metric1}
ds^2=\frac{1}{n^2} dt^2-dx^2 -dx_\perp^2.
\end{equation}

The photon creation process can be studied from the Lagrangian.  
\begin{eqnarray}
L_o&=&\frac{1}{2}\int dtdx^3  (\vec{E}\cdot \vec{D} -\vec{B}\cdot \vec{H} )\nonumber\\
&=&\frac{1}{2}\int dtdx^3  \Big(\epsilon (\partial_t \vec{A})^2 -(\nabla \times \vec{A} )^2 \Big),
\end{eqnarray}
here $\vec{E}= \partial_t \vec{A}$ and $B=\nabla \times \vec{A}$. Since this equation is valid only for one polarization, the probe field, $\vec{E}_s$, may be replaced by a scalar field $\phi_s$.  The effective Lagrangian is 
\begin{equation}
L_o \approx\frac{1}{2}\int dtdx^3  \Big(n^2( \partial_t \phi_s)^2 -(\nabla \phi_s )^2\Big) .
\end{equation}

\begin{figure}[h]
\includegraphics[width=4cm]{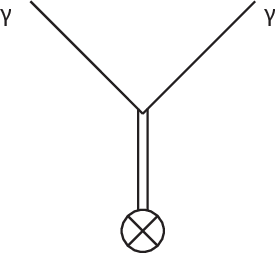}
\caption{ Two photons are created by the background fluctuation. The cross circle represents the background field, while vertical lines represent two background electric fields which then become two real photons.  
}
\label{vertice}
\end{figure}

When the high intensity pump field is inputted in a material whose background refractive index is $n_0=\sqrt{1+\chi^{(1)}}$, the free particle Lagrangian is 
\begin{equation}
L_f \approx\frac{1}{2}\int dtdx^3 \Big( n_0^2 (\partial_t \phi_s)^2 -(\nabla \phi_s )^2\Big) ,
\end{equation}
and the interaction Hamiltonian is 
\begin{equation}
H_I \approx  \xi \Pi_s^2  .
\end{equation}
Here,  $\xi=\frac{1}{2}(\frac{1}{n^2} -\frac{1}{n_0^{2}}$). $\Pi_s=n_0^2 \partial_t\phi$ is the conjugate momentum.
This interaction term represents two photons created by a background field which is converting two virtual photons to real photons from the background (fig. \ref{vertice}). These photon pairs are highly correlated.  

The free particle quantization is \cite{Hillery:1983sm,Drummond:1990ch,Glauber:1990ch,Schmidt:1996pk,Dalton:1996,Schutzhold:1998mu,Belgiorno:2014eja,2020JMOp...67..196R},  
\begin{eqnarray}
\label{qf1}
\phi_s &=&\int \frac{dk^3}{(2\pi)^3} \frac{1}{\sqrt{2\omega_{\vec{k}}}}  (a_{\vec{k}} e^{i\vec{k}\cdot \vec{r}}+a_{\vec{k}}^\dagger e^{-i\vec{k}\cdot \vec{r}}),\\
\label{qf2}
\Pi_s &=&-i\int \frac{dk^3}{(2\pi)^3} \sqrt{\frac{\omega_{\vec{k}}}{2}}  (a_{\vec{k}} e^{ i\vec{k}\cdot \vec{r}}-a_{\vec{k}}^\dagger e^{-i\vec{k}\cdot \vec{r}}),
\end{eqnarray}
here,
$\omega_{\vec{k}}=\frac{k}{n_0}$,
and
\begin{equation}
[a_{\vec{k}}, a_{\vec{k}'}^\dagger]=(2\pi)^3 \delta(\vec{k}-\vec{k}').
\end{equation}

It is well known that a strong field can generate particles. The
particle creation amplitude
\cite{Hillery:1983sm,Drummond:1990ch,Glauber:1990ch,Schmidt:1996pk,Dalton:1996,Schutzhold:1998mu,Belgiorno:2014eja,2020JMOp...67..196R}
is given by 
\begin{eqnarray}
A_{\vec{k} ;\vec{k}'}&=&\left\langle \vec{k} ;\vec{k}' \left\vert \exp(-i \int H_I dx^4) \right\vert 0 \right\rangle\nonumber\\
&\approx& -i \omega_k \omega_{k'}\tilde{\xi}(k+k'),
\end{eqnarray}
where, 
\begin{equation}
\label{Fxi}
\tilde{\xi}(k)=\int \xi(x,t)\exp(-ik x)dx^4.
\end{equation}

The created photon number is \cite{Hillery:1983sm,Drummond:1990ch,Glauber:1990ch,Schmidt:1996pk,Dalton:1996,Schutzhold:1998mu,Belgiorno:2014eja,2020JMOp...67..196R}
\begin{equation}
\label{pcreate}
N=\int \frac{dk^3}{(2\pi)^3} \frac{dk'^3}{(2\pi)^3}  \omega_k \omega_{k'}|\tilde{\xi}(k+k')|^2.
\end{equation}

Only one polarization mode is included in the calculation, which is different from the literature \cite{Hillery:1983sm,Drummond:1990ch,Glauber:1990ch,Schmidt:1996pk,Dalton:1996,Schutzhold:1998mu,Belgiorno:2014eja,2020JMOp...67..196R}.

\section{An expanding-space-like metric}

\begin{figure}[h]
\includegraphics[width=8cm]{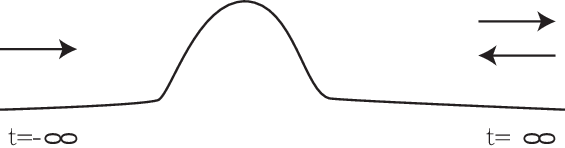}
\caption{  The refractive index changing with time: The black curve represents that the reflective index is a constant at $t\rightarrow -\infty$. It varies for a while, settles down, and returns to the original value. The left arrow represents a positive energy wave at the beginning. It then becomes a negative energy and positive energy in the future. This process induces particle creation. }
\label{refraction}
\end{figure}

Suppose the refractive index is a function of time only, $n(t)$. The corresponding analog metric is that of an expanding universe (characterized by some time-dependent scale factor $a(t)$). In this case, particles are also generated. The corresponding metric is chosen to be  
\begin{equation}
d\tau^2 = \frac{1}{n^3}dt^2 -\frac{1}{n}d\vec{x}^2,
\end{equation}
which gives the same equation of motion as eq. (\ref{eom}) and cannot be re-scaled to a flat space. This is a model giving by Parker and Toms\cite{Parker1,Parker2,Ford:2021syk} , in which the cosmological scale factor $a=n^{-1/2}$.  The refraction index, $n(t)$, is considered to be $n_0$ as $t\rightarrow \pm \infty$ for a relevant optical system (fig.\ref{refraction}).

The scalar field can be redefined as 
\begin{equation}
\phi_k= \frac{e^{i\vec{k}\cdot \vec{x}}}{(2\pi)^\frac{3}{2}}\chi_k(t)
\end{equation}
Then, the equation of motion reduces to 1-dimensional Schroding's equation form 
\begin{equation} 
-\partial_t^2 \chi_k + U(t) \chi_k= \omega^2 \chi_k
\end{equation} 
here, $U(t)= k^2 (\frac{1}{n_0^2}-\frac{1}{n^2})$ and $\omega=k/n_0$. If $\chi_k$ is chosen to be
\begin{equation}
\chi_k^{(in)}\sim \frac{e^{-i\omega_{in}t}}{\sqrt{2\omega_{in}}} \text{ , }t \rightarrow -\infty
\end{equation} 
The field is quantized in the form, 
\begin{equation} 
\hat{\phi} = \int \frac{dk^3}{(2\pi)^3\sqrt{2\omega_{in}}}  a_k \chi_k^{(in)} e^{i\vec{k}\cdot \vec{x}}+a^\dagger_k \chi_k^{(in)*} e^{i\vec{k}\cdot \vec{x}}
\end{equation}
 here, $\omega_{in}=k/n_0$ and $[a_k, a_k'^\dagger]=\delta(\vec{k}-\vec{k}')$. The $in$ vacuum is defined as. 
\begin{equation}
a_k \ket{in;0}=0
\end{equation}

One the other hand, if $\chi$ is chosen to be

 \begin{equation}
\chi_k^{(out)}\sim \frac{e^{-i\omega_{out}t}}{\sqrt{2\omega_{out}}} \text{ , }t \rightarrow \infty ,
\end{equation} 
where, 
 \begin{equation} 
\hat{\phi} = \int \frac{dk^3}{(2\pi)^3\sqrt{2\omega}}  b_k e^{-i\omega t} e^{i\vec{k}\cdot \vec{x}}+b^\dagger_k e^{i\omega t} e^{i\vec{k}\cdot \vec{x}} ,
\end{equation}
with $\omega_{out}=k/n_0$ and $[b_k, b_k'^\dagger]=\delta(\vec{k}-\vec{k}')$. The $out$ vacuum is defined as

\begin{equation}
b_k \ket{out;0}=0
\end{equation}
The required quantization condition is 
\begin{equation}
\chi_k \frac{d\chi^*_k}{dt}-\chi_k^* \frac{d\chi_k}{dt}=i
\end{equation} 
The Bogoliubov transformation is applied to study the particle creation. 
\begin{equation}
\chi^{(in)}_k \rightarrow \alpha_k \chi^{out}_k +\beta_k \chi_k^{out*}
\end{equation}
Based on the perturbation method\cite{Zeldovich:1971mw,Zeldovich:1977vgo,Birrell:1979pi}, the coefficients are 

\begin{eqnarray}
\alpha_k &\approx&  (1-\frac{i}{2\omega_{in}} \int^\infty_{-\infty} Udt)\\
\beta_k &\approx& \frac{i }{2 \omega_{in}} \int^\infty_{-\infty} U e^{-2i\omega_{in} t}dt= \frac{i \hat{ U} (2\omega_{in})}{2 \omega_{in}}
\end{eqnarray}
here, $\hat{U}(\omega)=\int^\infty_{-\infty} \xi(t) e^{-i\omega t}dt$ is the Fourier tranform of $U(t)$. 

 From the Bogolyubov transformation, the number density of the created particles are 

\begin{eqnarray}
N_e&=&\frac{1}{(2\pi )^3} \int d^3k |\beta_k|^2\nonumber\\
\label{eout}
&=&\frac{1}{(2\pi)^3 } \int d^3k \frac{|\hat{U}(2\omega_{in})^2|}{4\omega_{in}^2}
\end{eqnarray}

\section{Experimental setup to demonstrate  particle creation in an expanding universe}

It has been pointed out earlier that if an electric pulse varies in time much faster than in space, the resulting equation might resemble an expanding universe in a small region\cite{Schutzhold:2011ze}. Here, we demonstrate that two colliding laser beams have a component that satisfies this requirement. 
\begin{figure}[h]
\includegraphics[width=8cm]{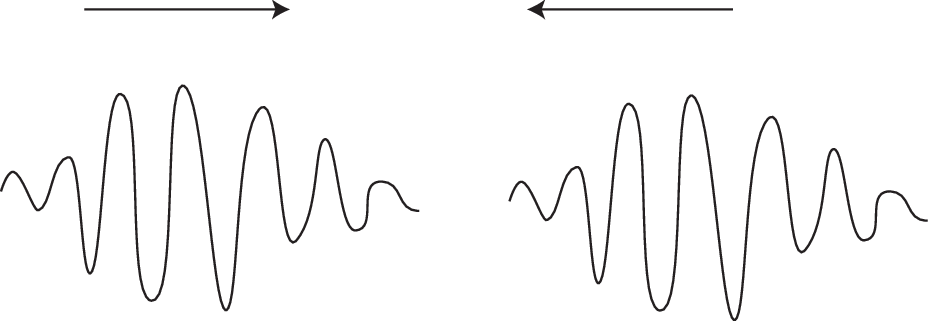}
\caption{ Two cos-wave laser beams which collide head to head. 
}
\label{collision}
\end{figure}

Consider two laser beams with electric fields $E_1$ and $E_2$ as  
\begin{equation}
E_i = A_i \exp\left[-\frac{\rho^2}{2\sigma_i^2}-\frac{(x\pm v_g t)^2}{2\sigma_g^2}\right]\cos(k_0 x \pm \omega_0 t).
\end{equation}

$E_1$ and $E_2$ are left and right moving waves, respectively. $\sigma_i$ is the width of the Gaussian wavepacket in the y
and z direction, and  $\sigma_g$ is the width of the Gaussian
wavepacket in the propagating direction (x
direction). $\rho^2=y^2+z^2$. The main angular frequency and wave
number are $\omega_0$ and $k_0$. We take   $  k_0\sigma_g\gg 1$ and $ k_0\sigma_i\gg 1$ to satisfy the expanding-universe-like requirement. The particles are created by the fluctuation of the reflective index, $\xi\approx -\frac{\chi^{(3)}}{2(1+\chi^{(1)})^2}|E|^2$. The fluctuation of the refractive index is proportional to the square of the electric field. When the laser beams collide, the electric field is in the superposition
\begin{equation}
(E_1+E_2)^2=E_1^2+2E_1 E_2+E_2^2,
\end{equation}
where the polarization directions of the two beams are the same. The first and third terms on the right side do not contribute to radiation if the motion is subluminal \cite{Belgiorno:2010zz,DallaPiazza:2012aj}. If the system is superluminal, it may be related to Cerenkov radiation or Hawking-like radiation. It is highly concentrated in a particular direction, which is not what we focus on right now. We focus on the second term or the cross-term. The cross term is 

\begin{equation}
2E_1 E_2= A_1 A_2
e^{-\frac{\rho^2}{\sigma_i^2}-\frac{x^2}{\sigma_g^2}-\frac{(v_g
    t)^2}{\sigma_g^2}} \left[\cos(2k_0 x)+\cos(2\omega_o t)
  \right].
\end{equation}

Based on eq. (\ref{Fxi}), the Fourier transform for term with $\cos(2k_o x)$ is 
\begin{equation} 
\tilde{E}_{12c}=A_1 A_2 \sigma_i^2 \sigma_g^2 v_g^{-1} \pi^2  e^{-\frac{k_\rho^2 \sigma_i^2}{4}}  e^{  -\frac{\left[(2k_0 \pm k)^2 +\frac{\omega^2}{v_g^2}\right]\sigma_g^2}{4}} .
\end{equation}
This term generates particles colinear to laser beams. These particles have the same energy as the laser beams. It is challenging to distinguish them from the background. Therefore, it is also not what we are interested in. 

Based on eq. (\ref{Fxi}), the Fourier transform for term with $\cos(2\omega_o t)$ is 
\begin{equation} 
\tilde{E}_{12s}=A_1 A_2 \sigma_i^2 \sigma_g^2 v_g^{-1} \pi^2  e^{-\frac{k_\rho^2 \sigma_i^2}{4}} e^{-\frac{\left[\frac{(2\omega_0 \pm \omega )^2}{v_g^2}+k^2\right] \sigma_g^2}{4} }.
\end{equation}

The term emits photon pairs in all directions. The photon pairs are created with energy of about $w_0$, and propagate in the opposite directions. They behave like scattered photons and can be separated from the background since the radiation is in all directions.

The total number of emitted photons  is \cite{Hillery:1983sm,Drummond:1990ch,Glauber:1990ch,Schmidt:1996pk,Dalton:1996,Schutzhold:1998mu,Belgiorno:2014eja,2020JMOp...67..196R}.
\begin{eqnarray}
N&=&\frac{I_1I_2n_1^2 \sigma_i^4\sigma_g^4\pi^4}{n_0^6 v_g^2}\int
     \frac{\omega_{k_1}\omega_{k_2}dk_1^3dk_2^3}{(2\pi)^6}
     e^{-\frac{k_\rho^2 \sigma_i^2+(\frac{(2\omega_0 \pm \omega )^2}{v_g^2}+k^2) \sigma_g^2}{2} }\nonumber\\
&\approx&  \frac{I_1 I_2n_1^2\sigma_i^2\pi}{4n_0^6\sigma_g^2} 
          (k_0\sigma_g)^4,
\end{eqnarray}
here, $\chi^{(3)}E_i^2=2n_0 n_1 I_i$. $\vec{k}=\vec{k}_1+\vec{k}_2$ and $\omega=\omega_{k_1}+\omega_{k_2}$. $I_i$ is intensity. Quartz's
refractive index related to the intensity is $n=n_0+n_1 I=1.5+3\times
10^{-16} I[cm^2/W]$. Consider a laser with wavelength  $\lambda=800\ nm$ in vacuum
 and waist $\sigma_i =10\ \mu m$. If the pulse length is about the same
as the width, $\sigma_g=\sigma_i$ and intensity is $10^{13}
\ W/cm^2$,  then the photon number is about 120 per pulse, compared with the
  total photon number of about $2\times 10^{12}$ per pulse in the pump
  laser beams.
  Even though the signal photon pairs have the same frequency as the pump laser's, they have an almost isotropic angular distribution different from the laser beams. The signal-to-noise ratio with order $10^{10}$ is achievable when detecting them at angles away from the original lasers.
  In a real experiment, the two colliding lasers may be aligned to a
  angle not equal to 180$^{\circ}$ to avoid damage of the optical parts.

\section{Conclusion}

It is widely accepted that a fluid dynamical system can exhibit analogous physical quantities to a system in curved space. Additionally, it is possible to construct a black-hole-like object and observe Hawking-like radiation in such systems. We choose a different matrix representing an expanding universe and keep the equation of motion the same as that of particle creation as the refractive index varies. Two head-to-head collision laser beams can construct the refractive index variance. This collision creates a small region where the refractive index varies very quickly with time. The refractive index variance then represents the universe expansion in the collision region. We demonstrate that the requirement is reachable under equipment precision and can be preceded in the laboratory.

\begin{acknowledgments}
D.C. Dai is supported by the National Science and Technology Council (under grant no. 111-2112-M-259-016-MY3)
C.B. Fu is supported by the National Nature Science Foundation of China (under grant no. 12235003)
\end{acknowledgments}

\end{document}